\documentclass[12pt]{article}
\usepackage{epsfig,amsmath,amssymb,latexsym,graphicx,setspace,fullpage}
\usepackage{chicago}
\usepackage{rotating}

\newcommand{\BR}{{\mathbb R}}

\usepackage{color} 

\begin{document}

\renewcommand{\thefootnote}{\fnsymbol{footnote}}

\title{Approximate Bayesian computation and \\ \vspace{0.1in}
Bayes linear analysis: Towards high-dimensional ABC} 

\author{David J. Nott\footnote{Department of Statistics and Applied Probability,
National University of Singapore, Singapore 117546.}, \:
Y. Fan\footnote{School of Mathematics and Statistics, University of New South Wales, Sydney 2052 Australia.}, 
\: L. Marshall\footnote{Department of Land Resources and Environmental Sciences,
Montana State University, 334 Leon Johnson Hall, PO Box 173120, Bozeman, MT 59717-3120} \: and\, S. A. Sisson$^\ddagger$\footnote{Communicating Author: Email \tt{Scott.Sisson@unsw.edu.au}}
}

\maketitle

\doublespacing

\vspace*{-8mm}\noindent

\begin{abstract}
\noindent Bayes linear analysis and approximate Bayesian computation (ABC) are techniques 
commonly used in the Bayesian analysis of complex models.  In this article
we connect these ideas by demonstrating that regression-adjustment ABC
algorithms produce samples for which first and second order moment summaries approximate
adjusted expectation and variance for a Bayes linear analysis.  
This gives regression-adjustment methods a useful interpretation and role in exploratory analysis 
in high-dimensional problems.  
As a result, we propose a new method for combining high-dimensional, 
regression-adjustment ABC with lower-dimensional approaches (such as using MCMC for ABC).
This method first obtains a rough estimate of the joint posterior via regression-adjustment ABC, 
and then estimates each univariate marginal posterior distribution separately in a lower-dimensional analysis. 
The marginal distributions of the initial 
estimate are then modified to equal the separately estimated marginals, thereby providing
an improved estimate of the joint posterior.    
We illustrate this method with several examples.
Supplementary materials for this article are available online.

\vspace{2mm}

\end{abstract}

\section{Introduction}

Bayes linear analysis and approximate Bayesian computation (ABC)  are two 
tools that have been widely used for the approximate Bayesian analysis of complex models.  
Bayes linear analysis can be thought of either as an approximation to a conventional
Bayesian analysis using linear estimators of parameters, or as a fundamental extension of 
the subjective Bayesian approach, where expectation rather than probability 
is a primitive quantity and only elicitation
of first and second order moments of variables of interest is required
(see e.g. \citeNP{goldstein+w07} for an introduction).  In this article, we are
interested in Bayes linear methods to approximate a conventional Bayesian analysis
based on a probability model, and in particular in the setting where
the likelihood is difficult to calculate.  We write $p(\theta)$ for the prior on
a parameter $\theta=(\theta_1,...,\theta_p)^{\top}$, $p(y|\theta)$ for the likelihood and $p(\theta|y)$ for
the posterior.  We discuss Bayes linear estimation further in the next section.  

Approximate Bayesian computation refers to a collection of methods which aim 
to draw samples from an approximate posterior distribution when the likelihood, 
$p(y|\theta)$, is unavailable or computationally intractable, but where it is feasible 
to quickly generate data from the model $y^*\sim p(y|\theta)$ 
(e.g. \shortciteNP{lopes+b09,bertorelle+bm10,beaumont10,csillery+bgf10,sisson+f11}).
The true posterior is approximated by $p(\theta|y)\approx p(\theta|s)$ where 
$s=s(y)=(s_1,\ldots,s_d)^{\top}$ is a low-dimensional vector of summary statistics 
(e.g. \shortciteNP{blum+ns11}). Writing
\begin{equation}
\label{eqn:abc-joint-post}
  p(\theta, s^*|s)\propto K_\epsilon(\|s-s^*\|)p(s^*|\theta)p(\theta),
\end{equation}
where $K_\epsilon(\|u\|)=K(\|u\|/\epsilon)/\epsilon$ is a standard smoothing kernel with scale 
parameter $\epsilon>0$, the approximate posterior itself is constructed as 
$p(\theta|s)\approx\int p(\theta,s^*|s) ds^*$, following standard kernel density estimation arguments.
The form of (\ref{eqn:abc-joint-post}) allows sampler-based ABC algorithms (e.g. 
\shortciteNP{marjoram+mpt03,bortot+cs07,sisson+ft07,toni+wsis09,beaumont+rmc09,drovandi+p11}) 
to sample from $p(\theta,s^*|s)$ without direct evaluation of the likelihood.

Regression has been proposed as a way to improve upon the conditional density estimation of 
$p(\theta|s)$ within the ABC framework. Based on a sample $(\theta^1,y^1),\ldots,(\theta^n,y^n)$ from 
$p(y|\theta)p(\theta)$, and then transforming this to a sample $(\theta^1,s^1),\ldots,(\theta^n,s^n)$ 
from $p(s|\theta)p(\theta)$ through $s^i=s(y^i)$, \shortciteN{beaumont+zb02} considered the 
weighted linear regression model
\begin{eqnarray}
\label{ramodel}
  \theta^i = \alpha+\beta^{\top}(s^i-s)+\varepsilon_i
  \qquad
  \mbox{for }i=1,\ldots,n,
\end{eqnarray}
where $\varepsilon_i$ are independent identically distributed errors, 
$\beta$ is a $d\times p$ matrix of regression coefficients and $\alpha$ is a $p\times 1$ vector. 
The weight for the pair $(\theta^i,s^i)$ is given by $K_\epsilon(\|s^i-s\|)$. This regression 
model gives a conditional density estimate of $p(\theta|s^i)$ for any $s^i$. For the observed 
$s$, this density estimate is an estimate of the posterior of interest, $p(\theta|s)$, and 
$\alpha+\varepsilon_i$ 
is a sample from it. Writing least squares estimates of $\alpha$ and $\beta$ 
as $\hat{\alpha}$ and $\hat{\beta}$, and the resulting empirical residuals as 
$\hat{\varepsilon}_i$, then the regression-adjusted vector
\begin{equation}
\label{linadjust}
  \theta^{i,a}=\theta^i-\hat{\beta}^{\top}(s^i-s)\approx \hat{\alpha}+\hat{\varepsilon}_i
\end{equation}
is approximately a draw from $p(\theta|s)$. \shortciteN{beaumont+zb02} do not consider the 
model (\ref{ramodel}) as holding globally, but instead consider a local-linear fit 
(this is expressed through specifying a kernel, $K_\epsilon$, with finite support). 
Variations on this idea include extensions to  generalised linear models 
\shortcite{leuenberger+w10} and non-linear, heteroscedastic regression based on a 
feed-forward neural network \shortcite{blum+f10}. The relative performance of the 
different regression adjustments are considered from a non-parametric perspective 
by \citeN{blum10}. However, application of regression-adjustment methods can fail 
in practice if the adopted regression model is clearly wrong, such as adopting the 
linear model (\ref{ramodel}) for a mixture, or mixture of regressions model.

The quality of the approximation $p(\theta|y)\approx p(\theta|s)$ depends crucially 
on the form of the summary statistics, $s$. Equality $p(\theta|y)=p(\theta|s)$ 
only occurs if $s$ is sufficient for $\theta$. However, reliably obtaining sufficient 
statistics for complex models is challenging \shortcite{blum+ns11}, and so an obvious 
strategy is to increase the dimension of the summary vector, $d=\dim(s)$, to include 
as much information about $y$ as possible. However, the quality of the second 
approximation, $p(\theta|s)\approx\int p(\theta,s^*|s) ds^*$, is largely dependent 
on the matching of vectors of summary statistics within the kernel $K_\epsilon$, 
which is itself dependent on the value of $\epsilon$. Through standard curse of 
dimensionality arguments (e.g. \citeNP{blum10}), for a given computational overhead 
(e.g. for a fixed number of samples $(\theta^i,s^i)$), 
the quality of the second approximation will 
deteriorate as $d$ increases. As a result, given that   more model parameters, 
$\theta$, imply more summary statistics, $s$, this reality is a primary reason 
why ABC methods have not, to date, found application in moderate to high-dimensional analyses.

In this article we make two primary contributions.  
First, we show there is an interesting
connection between Bayes linear analysis and regression-adjustment ABC methods.  
In particular, samples from the regression-adjustment ABC algorithm of 
\shortciteN{beaumont+zb02} result in first and second
order moment summaries which directly approximate Bayes linear adjusted expectation and
variance.  This gives the linear regression-adjustment method a useful interpretation for exploratory analysis 
in high dimensional problems.  

Motivated by this connection, our second contribution is to propose a new 
method for combining high-dimensional,
regression-adjustment ABC with lower-dimensional approaches, such as MCMC. 
This method first obtains a rough estimate of the joint posterior, $p(\theta|s)$, via 
regression-adjustment ABC, and then estimates each univariate marginal posterior 
distribution, $p(\theta_i|s)$, separately with a lower-dimensional
ABC analysis.  Estimation of marginal distributions is substantially easier than estimation of the joint 
distribution because of the lower dimensionality.  The marginal distributions of the initial 
estimate are then modified to be those of the estimated univariate marginals, thereby providing
an improved estimate of the joint posterior.  Similar ideas have been explored in the
density estimation literature (e.g. \shortciteNP{spiegelman+p03,hall+n06,giordani+mk09}).  
As a result, we are able to extend the applicability of ABC methods to problems of moderate 
to high dimensionality -- comfortably beyond current ABC practice.

This article is structured as follows: Section \ref{section:connection} 
introduces Bayes linear analysis, and explains its connection with
the regression-adjustment ABC method of \shortciteN{beaumont+zb02}.  Section \ref{section:method} describes
our proposed marginal adjustment method for improving the estimate of the ABC joint posterior distribution
obtained using regression-adjustment ABC.  A simulation study and real data analyses are presented in Section \ref{section:examples}, and Section \ref{section:discussion} concludes with a discussion.

\section{A connection between Bayes linear analysis and ABC}
\label{section:connection}

\subsection{Bayes linear analysis}
\label{section:connection1}

As in Section 1, suppose that $s=s(y)=(s_1,...,s_d)^{\top}$ is some vector of summary statistics based on data $y$, and
that $\theta=(\theta_1,...,\theta_p)^{\top}$ denotes parameter unknowns that we wish to learn about.  One view
of Bayes linear analysis (e.g. \citeNP{goldstein+w07}) is that it aims to construct an optimal linear 
estimator of $\theta$ under squared error loss. That is, an estimator of the form $a+Bs$, 
for a $p$-dimensional vector, $a$, and a $p\times d$ matrix, $B$, minimising
\[
  E[(\theta-a-Bs)^{\top}(\theta-a-Bs)].
\]
The optimal linear estimator is given by
\begin{eqnarray}
 E_s(\theta) & = & E(\theta)+\mbox{Cov}(\theta,s)\mbox{Var}(s)^{-1}[s-E(s)], \label{bayeslincrit}
\end{eqnarray}
where expectations and covariances on the right hand side are with respect to the joint
prior distribution of $s$ and $\theta$ i.e. $p(s|\theta)p(\theta)$. The estimator, 
$E_s(\theta)$, is referred to as the adjusted
expectation of $\theta$ given $s$.  If the posterior mean is a linear function of $s$
then the adjusted expectation and posterior mean coincide.  
Note that obtaining the best linear estimator of $\theta$ does
not require specification of a full prior or likelihood -- only first and second order 
moments of $(\theta,s)$ are needed.
From a subjective Bayesian perspective, the need to make only a limited
number of judgements concerning prior moments is a key advantage of the Bayes linear approach. 
There are various interpretations of Bayes linear methods -- see
\citeN{goldstein+w07}, for further discussion.  In the ABC context, a full probability model is typically available.
As such, we will consider Bayes linear analysis from a 
conventional Bayesian point of view as a computational approximation to a full Bayesian analysis.

The adjusted variance of $\theta$ given $s$, 
$\mbox{Var}_s(\theta)=E[(\theta-E_s(\theta))^{\top}(\theta-E_s(\theta)]$, can be shown to be
\[
  \mbox{Var}_s(\theta)=\mbox{Var}(\theta)-\mbox{Cov}(\theta,s)\mbox{Var}(s)^{-1}\mbox{Cov}(s,\theta).
\]
Furthermore, the inequality $\mbox{Var}_s(\theta)\geq E[\mbox{Var}(\theta|s)]$ holds, 
where $A\geq C$ means that $A-C$ is non-negative definite, and
the outer expectation on the right hand side is with respect to the prior distribution for $s$, $p(s)$.  
This inequality indicates that $\mbox{Var}_s(\theta)$ is a generally conservative upper
bound on posterior variance, although it should be noted that $\mbox{Var}_s(\theta)$ does not
depend on $s$, whereas $\mbox{Var}(\theta|s)$ is fully conditional on the observed $s$.  If the posterior
mean is a linear function of $s$, then $\mbox{Var}_s(\theta)=E[\mbox{Var}(\theta|s)]$

\subsection{Bayes linear analysis and regression adjustment ABC}

It is relatively straightforward to link the regression-adjustment approach of \shortciteN{beaumont+zb02} 
with a Bayes linear analysis. However, note that \shortciteN{beaumont+zb02}
do not consider the model (\ref{ramodel}) as holding globally, but instead assume that it holds locally 
around the observed summary statistics, $s$. We discuss this point further below, but for the 
moment we assume that the unweighted linear model
(\ref{ramodel}) holds globally, after an appropriate choice of the summary statistics,
$s$.

The ordinary least squares estimate of $\beta$ under the linear model 
(\ref{ramodel}) is  $\hat{\beta}=\hat{\Sigma}(s)^{-1}\hat{\Sigma}(s,\theta)$,
where $\hat{\Sigma}(s)$ is the sample covariance of $s^1,...,s^n$ and 
$\hat{\Sigma}(s,\theta)$ is the sample cross covariance of the pairs $(s^i,\theta^i)$, $i=1,...,n$.  
For large $n$ (where $n$ is a quantity under direct user control in an ABC analysis), 
$\hat{\beta}$ is approximately $\beta=\mbox{Var}(s)^{-1}\mbox{Cov}(s,\theta)$, 
where $\mbox{Var}(s)$ and $\mbox{Cov}(s,\theta)$ are the corresponding population versions
of $\hat{\Sigma}(s)$ and $\hat{\Sigma}(s,\theta)$.  Let $i$ be fixed and consider the sequence
of random variables 
$\theta^{i,a}=\theta^i-\hat{\beta}^{\top}(s^i-s)$ as $n$ tends to infinity.  
Note that $\hat{\beta}$ is a function of $(\theta^j,s^j)$ for $j=1,...,n$ here.  
By Slutsky's theorem, if $\hat{\beta}$ is consistent for $\beta$ then $\theta^{i,a}$ will
converge in probability and in distribution to $\theta^i-\beta^{\top}(s^i-s)$ as $n\rightarrow\infty$.  
Then we can write 
\begin{eqnarray*}
  \lim_{n\rightarrow\infty} E(\theta^{i,a}) & = & E[\theta^i-\beta^{\top}(s^i-s)] \\
                & = & E(\theta)+\mbox{Cov}(\theta,s)\mbox{Var}(s)^{-1}[s-E(s)] \\
                & = & E_s(\theta).
\end{eqnarray*}
where the interchange of limits in the first line can be justified by applying the Skorohod representation 
theorem and then the dominated convergence theorem.

By a similar argument
\begin{eqnarray*}
  \lim_{n\rightarrow\infty} \mbox{Var}(\theta^{i,a}) & = & \mbox{Var}[\theta^i-\beta^{\top}(s^i-s)] \\
   & = & \mbox{Var}(\theta)+\beta^{\top}\mbox{Var}(s)\beta-2\mbox{Cov}(\theta,s)\beta \\
   & = & \mbox{Var}(\theta)-\mbox{Cov}(\theta,s)\mbox{Var}(s)^{-1}\mbox{Cov}(s,\theta) \\
   & = & \mbox{Var}_s(\theta).
\end{eqnarray*}
Hence, the covariance matrix of the regression-adjusted $\theta^{i,a}$ approximates the Bayes linear adjusted variance
for large $n$.  

These results demonstrate that the first and second moments of the regression-adjusted 
samples $\theta^{i,a}$, $i=1,...,n$
in the linear method of \shortciteN{beaumont+zb02} have a useful interpretation, 
regardless of whether the linear assumptions of the regression model (\ref{ramodel}) 
hold globally, as a Monte Carlo approximation to a Bayes linear analysis.  
This connection with Bayes linear analysis is not surprising when one considers that a Monte Carlo
approximation to (\ref{bayeslincrit}) based on draws from the prior is just a least squares
criterion for regression of $\theta$ on $s$. 
Usefully for our present purposes, the Bayes linear interpretation may be helpful for motivating an exploratory
use of regression adjustment ABC, even in problems of high dimension.  
In high-dimensional problems, an anonymous referee has suggested 
it might also be useful to consider more sophisticated shrinkage
estimates of covariance matrices in implementing the Bayes linear approach.  
The connection between Bayes linear methods and regression-adjustment ABC 
continues to hold if kernel weighting is reincorporated into the regression model (\ref{ramodel}).
Now consider the model (\ref{eqn:abc-joint-post}) in general and 
a Bayes linear analysis using first and second order moments of $(\theta,s^*)|s$ with
Bayes linear updating by the information $s=s^*$.  This then corresponds to the kernel
weighted version of the procedure of \shortciteN{beaumont+zb02}.  

A recent extension of regression-adjustment ABC is 
the nonlinear, heteroscedastic method of \citeN{blum+f10} which replaces (\ref{ramodel}) with
\begin{eqnarray}
  \theta^i & = & \mu(s^i)+\sigma(s^i)\varepsilon^i,  \label{nonlinearmodel}
\end{eqnarray}
where $\mu(s^i)=E(\theta|s=s^i)$ is a mean function, $\sigma(s^i)$ is a diagonal matrix
with diagonal entries equal to the square roots of the diagonal entries of $\mbox{Var}(\theta|s=s^i)$, 
and the $\varepsilon^i$ are i.i.d.
zero mean random vectors with $\mbox{Var}(\varepsilon^i)=I$. 
It is possible also to take $\sigma(s)$ to be some matrix square root of $\mbox{Var}(\theta|s=s^i)$
where all elements are functions of $s$.  
If (\ref{nonlinearmodel}) holds, then the adjustment
\[
  \theta^{i,a}=\mu(s)+\sigma(s)\sigma(s^i)^{-1}[\theta^i-\mu(s^i)]
\]
is a draw from $p(\theta|s)$.  The heteroscedastic adjustment approach does seem to be outside
the Bayes linear framework. However, a nonlinear mean model for $\mu(s)$ with a constant
model for $\sigma(s)$ can be reconciled with
the Bayes linear approach by considering an appropriate basis expansion involving functions of $s$.  
\citeN{blum10} gives some theoretical support for more complex regression adjustments through
an analysis of a certain quadratic regression adjustment and suggests that transformations of $\theta$
can be used to deal with heteroscedasticity.    
In this case, the Bayes linear interpretation would be more broadly applicable in regression-adjustment ABC.  
Another recent regression adjustment approach is
that of \citeN{bonassi11}, which is based on fitting a flexible mixture model to the joint samples.

An interesting recent related development is the semi-automatic method of choosing summary statistics
of \citeN{fernhead+p12}.  They consider an initial provisional set of statistics and then use
linear regression to construct a summary statistic for each parameter, based on samples from the prior
or some truncated version of it.  Their approach can be seen as a use of Bayes linear estimates as
summary statistics for an ABC analysis.  There are several other innovative aspects of their paper
but their approach to summary statistic construction provides another strong link with Bayes linear
ideas.

\section{A marginal adjustment strategy}
\label{section:method}

Conventional sampler-based ABC methods, such as MCMC and SMC, which use rejection- or importance weight-based 
strategies, are hard to apply in problems of moderate or
high dimension.  This occurs as an increase in the dimension of the parameter, $\theta$, 
forces an increase in the dimension of the summary statistic, $s$. This, in turn, causes performance 
problems for sampler-based ABC methods as the term $K_{\epsilon}(\|s-s^*\|)$ in 
(\ref{eqn:abc-joint-post}) suffers from the curse of dimensionality \cite{blum10}.
On the other hand, regression-adjustment strategies, which can often be interpreted as Bayes linear 
adjustments (see Section \ref{section:connection}),  can be useful in problems with many parameters.
However, it is difficult to validate their accuracy, and sampler-based ABC methods may be preferable in low dimensional
problems, particularly when simulation under the model is computationally inexpensive.

We now suggest a new approach to combining the low-dimensional accuracy of sampler-based ABC methods, 
with the utility of the higher-dimensional, regression-adjustment approach.
In essence, the idea is to construct a first rough estimate of the approximate posterior using 
regression-adjustment ABC, and also separate estimates of each of the marginal distributions of 
$\theta_1|s,\ldots, \theta_p|s$. Estimating marginal distributions is easier than the full posterior, 
because of the reduced dimensionality of summary statistics required to be informative about a single 
parameter. Because of the lower dimensionality, each marginal density can often be more precisely 
estimated by any sampler- or regression-based ABC method, than the same margin of the regression-based 
estimate of the joint distribution.
We then adjust the marginal distributions of the rough posterior estimate to be those of the separately
estimated marginals, by an appropriate replacement of order statistics.  
The adjustment of the marginal distributions maintains the
multivariate dependence structure in the original sample.
When the marginals are well estimated, it is reasonable to expect that the 
joint posterior is better estimated. 

Precisely, the procedure we use is as follows:
\begin{enumerate}
\item Generate a sample $(\theta^i,s^i)$, $i=1,...,n$ from $p(\theta)p(s|\theta)$.  
\item Obtain a regression adjusted sample $\theta^{i,a}$, $i=1,...,n$ based on either the
model (\ref{ramodel}) or (\ref{nonlinearmodel}) fitted to the sample generated at Step 1.  
The regression adjusted methods may be implemented with or without kernel weighting.  
\item For $j=1,...,p$, 

\begin{enumerate}
\item For the marginal model for $\theta_j$,
\[
  p(y|\theta_j)=\int p(y|\theta)p(\theta_{-j}|\theta_j)d\theta_{-j},
\]
where $\theta_{-j}$ is $\theta$ with the element $\theta_j$ excluded,
identify summary statistics $s(j)=(s(j)_1,...,s(j)_{d(j)})^{\top}$ that
are marginally informative for $\theta_j$.

\item Use a conventional ABC method to estimate the posterior distribution for
$\theta|s(j)$.  Extracting the $j^{th}$ component results in a sample,
$\theta_j^{m,i}$, $i=1,...,n$.  If the number of samples drawn is not $n$, 
then we obtain a density estimate based on the samples we have and then define
$\theta_j^{m,i}$, $i=1,...,n$ to be $n$ equally spaced quantiles from the density estimate.  

\item Replace the $i=1,\ldots, n$ order statistics for the $j^{th}$ component of the sample $\theta^{a,i}$, 
by the equivalent quantiles of the marginal samples $\theta_j^{m,i}$.

More precisely, writing $\theta_j^{m,(k)}$ and $\theta_j^{a,(k)}$ as the $k^{th}$ order statistic of the samples $\theta_j^{m,i}$ and $\theta_j^{a,i}$, respectively, $i=1,...,n$
then we replace $\theta_j^{a,(i)}$ with $\theta_j^{m,(i)}$ for $i=1,\ldots,n$.

\end{enumerate}
\end{enumerate}
The samples, $\theta^{a,i}$,  with all $p$ margins adjusted are then taken as an 
approximate sample from the ABC posterior distribution.  The samples used in Step 3 in the above
algorithm can either be the same as those generated in Step 1 or generated independently, but
to save computational cost we suggest using the same samples.  An anonymous referee has suggested
that it may be possible to make use of the adjusted joint samples to help choose the summary statistics
for the marginals, and this is an intriguing suggestion but something that we leave to future work.  
The same referee has also pointed out the danger that the estimated marginals might not be compatible
with the dependence structure in the joint distribution if there are parameter constraints - indeed, 
this can happen for ordinary regression adjustment approaches.  However, mostly such problems can
be dealt with through an appropriate reparameterisation.  

The idea of incorporating knowledge of marginal distributions into estimation of a joint distribution
has been previously explored in the density estimation literature.  \citeN{spiegelman+p03}
consider parametrically estimated marginal distributions and then replacing order statistics in
the data by the quantiles of the parametrically estimated marginals.  This is similar in spirit
to our adjustment procedure in the ABC context.  They show by theoretical arguments and examples
that improvements can be obtained if the parametric assumptions are correct.  \citeN{hall+n06}
consider density estimation when there is a dataset of the joint distribution as well as additional
datasets for the marginal distributions.  They consider a copula approach to estimation of the joint
density and show that the additional marginal information is beneficial if the copula is sufficiently
smooth.  Recently, \shortciteN{giordani+mk09} have considered a mixture of normals copula approach
where the marginals are also estimated as mixtures of normals.  

A powerful motivation for using available marginal information  comes from the fact that a joint
distribution is determined by the univariate distributions of all its linear projections.  
This arises as the characteristic function of the joint distribution
is determined from the characteristic functions of one dimensional projections \cite{cramer+w36}.  
Hence adjusting the distribution of all linear projections of a density estimate to be correct 
would result in the true distribution being obtained.  By adjusting marginal distributions we
only consider a selected small number of linear projections. However, we expect that if good
estimates of marginal distributions are available, then transforming a rough estimate of the joint
density to take these marginal distributions will be beneficial.  

Note that estimation and adjustment of the $p$ marginal distributions in Step 3 may 
be performed in parallel, so that computation scales well with the dimension of $\theta$. 
Because the Bayes linear adjusted variance, $\mbox{Var}_s(\theta)$, is generally a conservative 
upper bound on the posterior variance (see Section \ref{section:connection1}), it is credible 
that the initial rough samples $\theta^{i,a}$ could form the basis of initial sampling distributions 
for importance-type ABC algorithms (e.g. \shortciteNP{sisson+ft07,beaumont+rmc09,drovandi+p11}), 
resulting in potential computational savings.
Finally we note that a number of methods exist to quickly determine the appropriate statistics, 
$s(j)$, for each marginal analysis. The reader is referred to \shortciteN{blum+ns11} for a 
comparative review of these.

\section{Examples}
\label{section:examples}

\subsection{A Simulated Example}

We first construct a toy example where the likelihood can be evaluated and where a gold standard 
answer is available for comparison. While ABC methods are not needed for the analysis of this model, 
it is instructive for understanding the properties of our methods. 
We consider a $p$-dimensional Gaussian mixture model with $2^p$ mixture components. 
The likelihood for this model is given by
\[
p(s|\theta)=\sum_{b_1=0}^1 \cdots \sum_{b_p=0}^1 \left[\prod_{i=1}^p \omega^{1-b_i}(1-\omega)^{b_i}\right] \phi_p(s| \mu(b,\theta),\Sigma),
\]
where $\phi_p(x| a, B)$ denotes the $p$-dimensional Gaussian density function with mean 
$a$ and covariance $B$ evaluated at $x$, $\omega\in[0,1]$ is a mixture weight, 
$\mu(b,\theta)^\top=((1-2b_1)\theta_1,\ldots,(1-2b_p)\theta_p)$, $b=(b_1,\ldots,b_p)$ 
with $b_i\in\{0,1\}$ and $\Sigma=[\Sigma_{ij}]$ is such that $\Sigma_{ii}=1$ and 
$\Sigma_{ij}=\rho$ for $i\neq j$. Under this setting, the marginal distribution 
for $s_i$ is given by the two-component mixture
\begin{eqnarray}
\label{true-margin}
  p(s_i|\theta_i) = (1-\omega)\phi_1(s_i| -\theta_i,1) + \omega\phi_1(s_i|\theta_i,1).
\end{eqnarray}
The combination of the $p$ two-mixture-component marginal distributions forms the $2^p$ 
mixture components for the $p$-dimensional model. 
Given $\theta$, data generation under this model proceeds by independently
generating each component of $b$ to be $0$ or $1$ with probabilities $\omega$ and $1-\omega$ respectively, 
and then drawing
$s|\theta,b\sim \phi_p(\mu(b,\theta),\Sigma)$.

For the following analysis we specify $s=(5, 5, \ldots, 5)$, $\omega=0.3$ and  $\rho=0.7$, 
and restrict the posterior to have finite support on $[-20,40]^p$, over which we have a uniform 
prior for $\theta$. Computations are performed using 1 million simulations from $p(s|\theta)p(\theta)$, 
using a uniform kernel $K_\epsilon(\|u\|)$, where $\|\cdot\|$ denotes Euclidean distance, and 
where $\epsilon$ is chosen to select the 10,000 simulations closest to $s$. We contrast results 
obtained using standard rejection sampling, rejection sampling followed by the regression-adjustment 
of \shortciteN{beaumont+zb02}, and both of these after applying our marginal-adjustment strategy. 
All inferences were performed using the {\tt R} package {\tt abc} \shortcite{csillery+fb11}.

\begin{sidewaysfigure}
\begin{center}
\includegraphics[width=22cm]{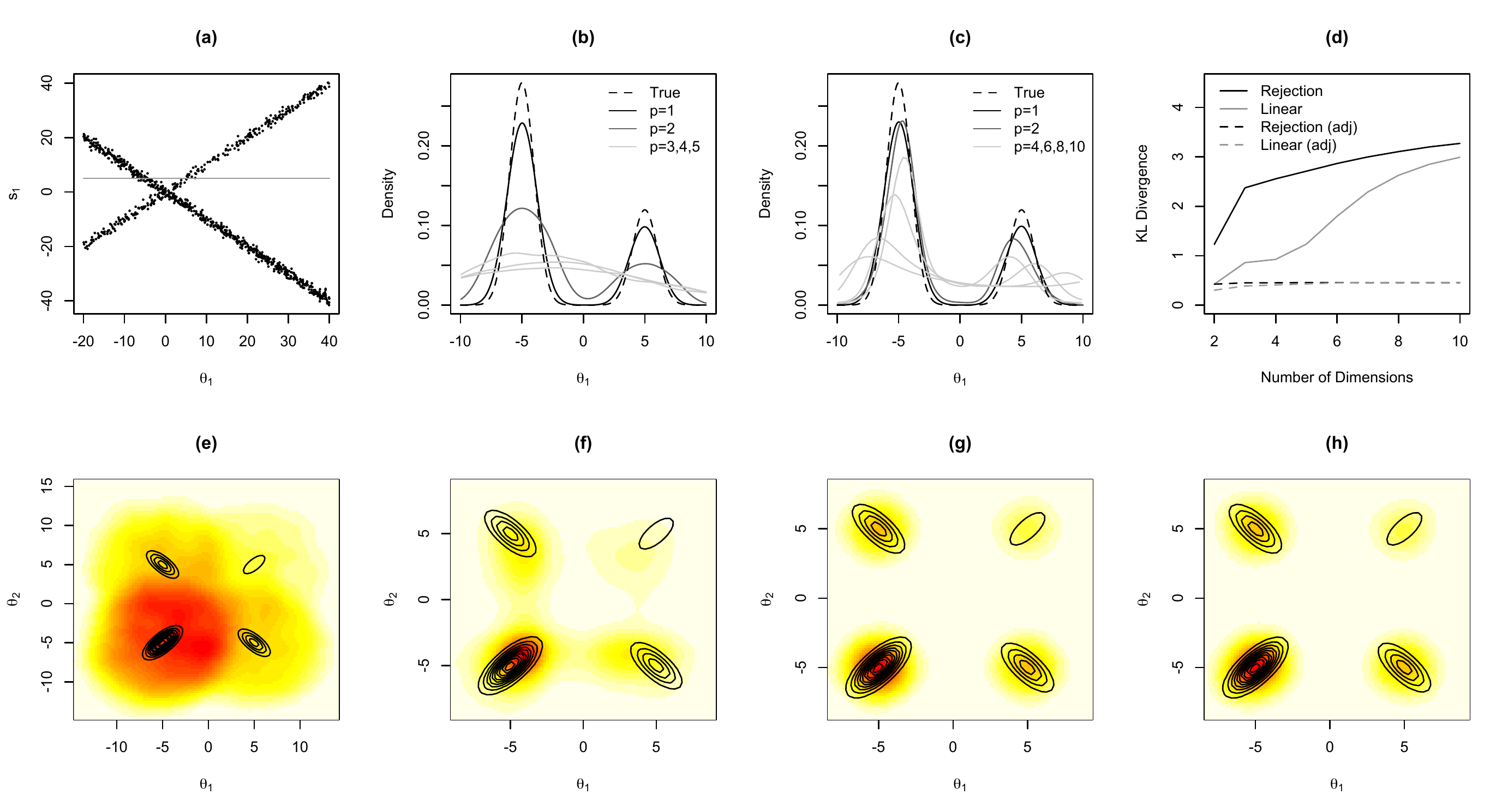}
\end{center}
\caption{ \label{mixture-plots} 
Inference on the Gaussian mixture model: Panel (a) plots the relationship between $\theta_1$ and $s_1$. 
The horizontal line corresponds to the observed summary statistic, $s_1=5$. Panels (b) and (c) show 
the estimated marginal posterior for $\theta_1$ respectively using rejection sampling, and rejection 
sampling followed by regression-adjustment. Results are shown for a range of model dimensions, $p$. 
True marginal posterior is indicated by the dashed line. Panel (d) shows the quality of the ABC 
approximation to the true model for the first two model parameters, $(\theta_1,\theta_2)$, as 
measured by the Kullback-Leibler divergence, as a function of model dimension, $p$. Black and grey 
lines respectively denote rejection sampling and rejection sampling followed by regression adjustment. 
Dashed lines indicate use of our marginal adjustment strategy. Panels (e)--(h) illustrate the ABC 
approximation to the bivariate posterior $(\theta_1,\theta_2)|s$ when $p=3$. Contours indicate the 
true model posterior. Panels (e) and (f) correspond respectively to rejection sampling, and 
rejection sampling followed by regression adjustment. Panels (g) and (h) correspond to panels 
(e) and (f) with the addition of our marginal adjustment strategy.}
\end{sidewaysfigure}

Figure \ref{mixture-plots}(a) illustrates the relationship between $\theta_1$ and $s_1$ 
(all margins are identical), with around 70\% of summary statistics located in the line 
with negative slope. The observed summary statistic is indicated by the horizontal line, 
the marginal posterior distribution for $\theta_1$ defined by the implied density of summary 
statistics on this line. Figure \ref{mixture-plots}(b) shows density estimates of $\theta_1|s$ 
using rejection sampling for $p=1, 2, \ldots, 5$ model dimensions. The univariate true marginal 
distribution is indicated by the dashed line. As model dimension increases, 
the quality of the approximation to the true marginal distribution deteriorates rapidly. This 
is due to the curse of dimensionality in ABC (e.g. \citeNP{blum10}) in which the restrictions 
on $s_1$ for a fixed number of accepted samples (in this case 10,000) decrease within the 
comparison $\|s-s^*\|$ as $p$ increases.  Of course one could increase
the number of simulations as the dimension increases, but assuming a fixed
computational budget returning a fixed number of samples provides one perspective
on the curse of dimensionality here.  Beyond $p=2$ dimensions, these density estimates 
are exceptionally poor. The same information is illustrated in Figure 
\ref{mixture-plots}(c) after applying the linear regression-adjustment of \shortciteN{beaumont+zb02} 
to the samples obtained by rejection sampling in Figure \ref{mixture-plots}(b). Clearly the 
regression-adjustment is beneficial in providing improved marginal density estimates. 
However, the quality of the approximation still deteriorates quickly as $p$ increases, 
albeit more slowly than for rejection sampling alone.

Figures \ref{mixture-plots}(e) and (f) show the two dimensional density estimates of $(\theta_1,\theta_2)|s$ 
for the $p=3$ dimensional model, respectively using rejection sampling, and rejection sampling 
followed by the linear regression-adjustment. The superimposed contours correspond to those of 
the true bivariate marginal distribution. The improvement to the density estimate following the 
regression-adjustment is clear, however even here, the component modes appear to be slightly 
misplaced, and there is some blurring of density with neighbouring components.

Figures \ref{mixture-plots}(g) and (h) correspond to the densities in Figures \ref{mixture-plots}(e) 
and (f) after the implementation of our marginal adjustment strategy. Here, each margin of the 
distributions is adjusted to be that of the appropriate univariate marginal density estimate in a 
$p=1$ dimensional analysis. E.g. the margins for $\theta_1$ are adjusted to be exactly the ($p=1$) 
density estimates in Figures \ref{mixture-plots}(b) and (c). In both plots (g and h) there is a 
clear improvement in the bivariate density estimate: the locations of the mixture components are in 
the correct places, and on the correct scales. Some of the accuracy of the dependence structure is 
less well captured under just rejection sampling, however (Figure \ref{mixture-plots}(g)). Here, the 
correlation structure of each Gaussian component seems to be poorly estimated, compared to that 
obtained under the regression-adjustment transformed samples. The message here is clear: any marginal 
adjustment strategy cannot recover from a poorly estimated dependence structure. The regression 
adjusted density in Figure \ref{mixture-plots}(f) more accurately captures the correlation structure 
of the true density, and this improved dependence structure is carried over to the final density 
estimate in Figure \ref{mixture-plots}(h).

Finally, Figure \ref{mixture-plots}(d) examines the quality of the ABC approximation to the true density, 
$p(\theta|s)$. Plotted is the Kullback-Leibler divergence, 
$\int p(\theta)\log\left[p(\theta)/q(\theta)\right]d\theta$, between the densities of the 
first two dimensions of each distribution as a function of model dimension (where $p(\theta)=p(\theta_1,\theta_2|s)$ is the true density, and $q(\theta)=q(\theta_1,\theta_2|s)$ 
is a kernel density estimate of the ABC approximation). The divergence is computed by Monte 
Carlo integration using 2,000 draws from the true density. We compare only the first two 
dimensions of the $p$-dimensional posteriors to maintain computational accuracy, 
noting that all pairwise marginal distributions of the full 
posterior are identical in this analysis (similarly for all higher-dimensional marginals) . 

Figure \ref{mixture-plots}(d) largely supports our previous conclusions. The performance of the 
rejection sampler and the rejection sampler with regression-adjustment deteriorates rapidly 
as the number of model dimensions (i.e. summary statistics) increases, although the latter 
performs better in this regard. There is a clear improvement to both of these approaches gained 
though our marginal adjustment strategy, with the modified regression-adjustment samples 
performing marginally better (for this example) where the original regression-adjustment 
provides better estimates of the multivariate dependence structure in lower dimensions.

After around $p=5$ dimensions there is little difference between the two marginally adjusted 
posteriors, and the divergence levels off to a constant value independent of model dimension. 
This is result of the ABC setup for this analysis. Beyond around $p=5$ dimensions, there is 
little difference between the rejection sampling and regression-adjusted posteriors 
(e.g. Figures \ref{mixture-plots}(e) and (f)), both largely representing near-uniform 
distributions over $\theta$. Hence, our marginal adjustment strategy is only able to make 
the same degree of improvements, regardless of model dimension.
The correlation dependence structure is also lost beyond this point, so the expected 
benefit of the regression-adjustment prior to marginal regression adjustment, is nullified. 
Using a lower initial threshold, $\epsilon$ (computation permitting), would allow a more 
accurate initial ABC analysis, and hence more discrimination between the rejection 
sampling and regression-adjustment approaches.

\subsection{Excursion set model for heather incidence data}

\begin{figure}
\begin{center}
\begin{tabular}{c}
\includegraphics[width=120mm,height=60mm]{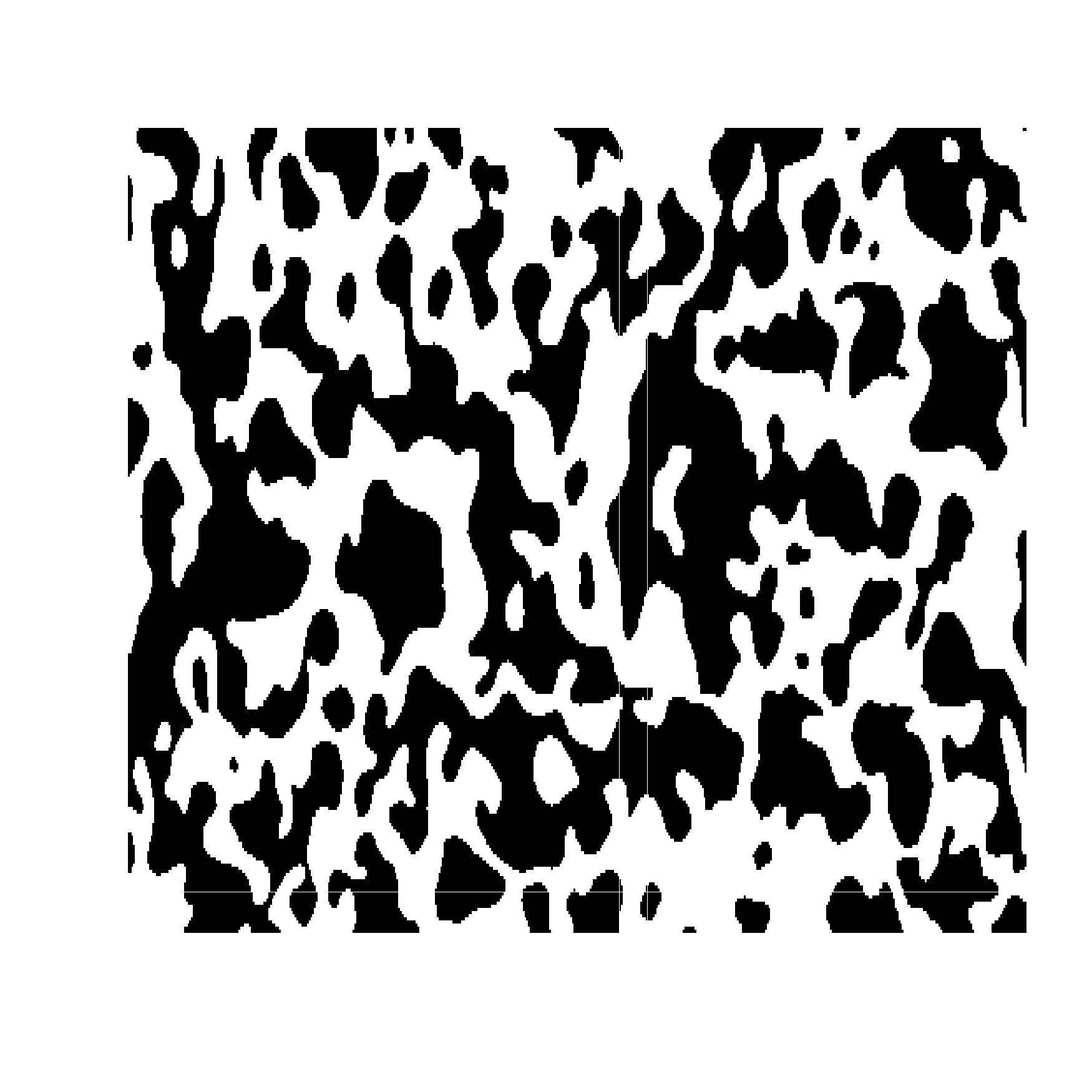} \\
\end{tabular}
\end{center}
\caption{\label{heather} 
The  heather incidence data representing a $10\times 20$ metre region (Diggle, 1991).}
\end{figure}

We now consider the medium resolution version of the heather incidence data 
analysed by \citeN{diggle81}, which is available in the {\tt R} package {\tt spatstat} \cite{baddeley+t05}.  
Figure \ref{heather} illustrates the data, consisting of a $256\times 512$ grid of zeros and ones, with each binary variable representing
presence (1) or absence (0) of heather at a particular spatial location. 
\citeN{nott+r99}
used excursion sets of Gaussian random fields to model a low resolution version
of these data.  Without loss of generality, we assume that the data are observed on an integer lattice.  

Let $\{Y(t); t\in \BR^2\}$ be a stationary Gaussian random field with mean zero and covariance function
\begin{eqnarray*}
 R(s,t) & = & \mbox{Cov}(Y(s),Y(t))=\exp[-(t-s)^{\top} A (t-s)]  \label{covfnmodel}
\end{eqnarray*}
where $s,t\in\BR^2$ and where $A$ is a symmetric positive definite matrix.  
Hence $R(s,t)$ corresponds to the Gaussian covariance function model with elliptical
anisotropy.  The $u$-level excursion set  of $Y(t)$ is defined as $E_u(Y)=\{t\in \BR^2: Y(t)\geq u\}$,
so that $E_u(Y)$ is obtained by ``thresholding" $Y(t)$ at level $u\in \BR$.  For
background on Gaussian random fields and geometric properties of their excursion sets
see e.g. \citeN{adler+t07}.

We model the heather data as binary random variables which are indicators for inclusion
in an excursion set on an integer lattice.  The data are denoted $B=\{B(i,j): i=0,...,255, j=0,...,511\}$
where $B(i,j)=I((i,j)\in E_u(Y))$ and where $I(\cdot)$ denotes the indicator function.  
The distribution of $B$ clearly depends on $u$ and $A$.  We write the $(i,j)$th element of
$A$ as $A_{ij}$.  Since $A$ is symmetric, $A_{12}=A_{21}$.   We parametrize the distribution
of $B$ through $\theta=(\theta_1,\theta_2,\theta_3,\theta_4)$ where $\theta_1=u$, 
$\theta_2=\log A_{11}$, $\theta_3=\log A_{22}$ and 
$\theta_4=\mbox{logit}[(A_{12}/\sqrt{A_{11} A_{22}}+1)/2]$.  We adopt the independent prior distributions
$\theta_1\sim N(0,0.5^2)$, 
$\theta_2,\theta_3\sim N(-4,0.5^2)$ and $\theta_4\sim N(0,0.5^2)$.  
The priors here are fairly noninformative.  For example, note that the expected fraction of
ones in the image is $1-\Phi(\theta_1)$.  Hence if we were to use a normal prior
for $\theta_1$ with very large variance, this would give roughly prior probability 0.5 to
having all zeros and probability 0.5 to have all ones.  This is an unreasonable and highly
informative prior.  Our prior is fairly noninformative for the expected fraction of ones.  
Similarly our prior on $\theta_2$ and $\theta_3$ assigns roughly 95\% prior probability
for correlation in the east-west and north-south directions at lag 10 respectively being between
zero and 0.5. For $\theta_4$, there is approximately 95\% prior probability that 
$A_{12}/\sqrt{A_{11} A_{22}}$ (a kind of correlation parameter describing the degree of
anistropy for the covariance function) lies between $-0.5$ and $0.5$.  
The results reported below are not sensitive to changes in these priors within reason, but
as mentioned highly diffuse priors with large variances are not appropriate.  
Simulation of Gaussian random fields is achieved with the {\tt RandomFields} package
in {\tt R} \cite{schlather11}, using the circulant embedding algorithm of
\citeN{dietrich+n93} and \citeN{wood+c94}.

For summary statistics,
denote by $n_{11}(v)$ for $v\in \BR^2$ the number of pairs of variables in $B$, separated by
displacement $v$, which are both 1.  Similarly denote by $n_{00}(v)$ the number
of such pairs which are both zero, and by $n_{01}(v)$ the number of
pairs where precisely one of the pair is 1 (the order does not matter).  
In terms of estimating each marginal distribution $\theta_1|s(1),\ldots,\theta_4|s(4)$,  we specify 
\begin{eqnarray*}
s(1)& = &
\sum_{i,j} B(i,j)/(256\times 512)\\
s(2)&=&[n_{11}(v_1),n_{00}(v_1),n_{01}(v_1)]^{\top}\\
s(3)&=&[n_{11}(v_2),n_{00}(v_2),n_{01}(v_2)]^{\top}\\
s(4)&=&[n_{11}(v_3),n_{11}(v_4),n_{00}(v_3),n_{00}(v_4),n_{01}(v_3),n_{01}(v_4)]^{\top}
\end{eqnarray*}
as the summary statistics for each parameter, 
where $v_1=(0,1)$, $v_2=(1,0)$, $v_3=(1,1)$ and $v_4=(1,-1)$. 

For the joint posterior regression-adjustment, we used the 
heteroscedastic, non-linear regression (\ref{nonlinearmodel}) \cite{blum+f10}, 
using the uniform kernel, $K_\epsilon(\|\cdot\|)$, with scale parameter set to 
give non-zero weight to all $2,000$ samples 
$(\theta^i,s^i)\sim p(s|\theta)p(\theta)$, and where $\|\cdot\|$ represents scaled 
Euclidean distance.  
The individual marginal distributions were estimated in the same manner, but with the 
kernel scale parameter specified to select the 1,000 simulations closest to each $s(j)$.
All analyses were again performed using the {\tt R} package {\tt abc} \shortcite{csillery+fb11}
with the default settings for the heteroscedastic nonlinear method.  

\begin{figure}
\begin{center}
\begin{tabular}{c}
\includegraphics[width=100mm]{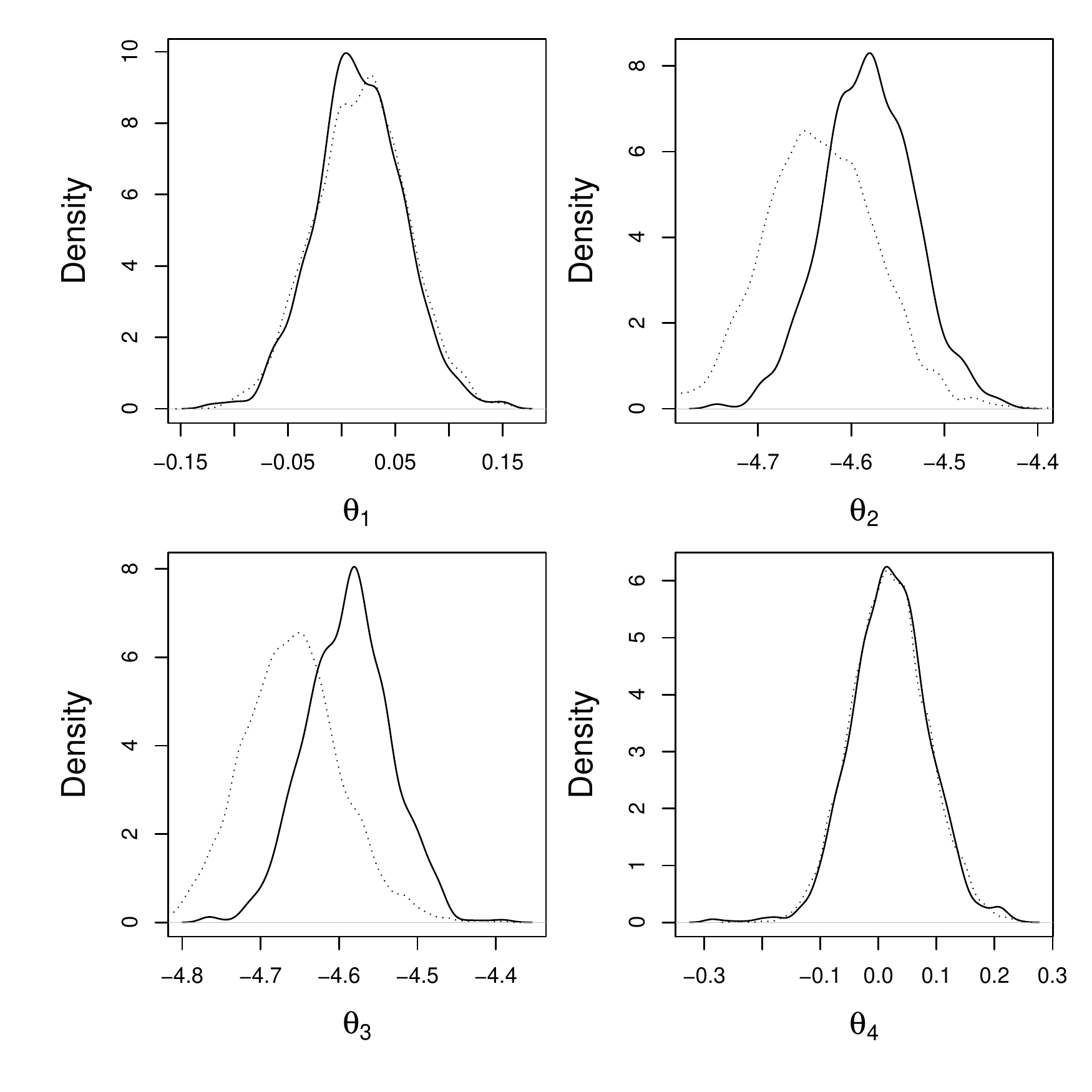} \\
\end{tabular}
\end{center}
\caption{\label{excmarginals} 
Estimated marginal posterior distributions of $\theta_1,\ldots,\theta_4$ 
for the heather incidence analysis.  Solid lines
denote individually estimated marginals; dotted lines illustrate estimated margins from the joint posterior analysis}
\end{figure} 

Figure \ref{excmarginals} shows estimated marginal posterior
distributions for the components of $\theta$ obtained by the joint regression-adjustment (dotted lines), 
and the same margins following our marginal adjustment strategy (solid lines).
The estimates for the spatial dependence parameters $\theta_2$ and $\theta_3$ are poor for
the joint regression approach -- the individually estimated marginals are estimated very accurately, which
can be verified by a rejection based analysis with a much larger number of samples (results not shown).  
Clearly if we use samples from
the approximate joint posterior distribution from the global regression 
for predictive inference or other purposes, the fact that the unadjusted
marginals are centred in the wrong place can lead to unacceptable performance of the approximation.  

\begin{figure}
\begin{center}
\begin{tabular}{cc}
\includegraphics[width=70mm]{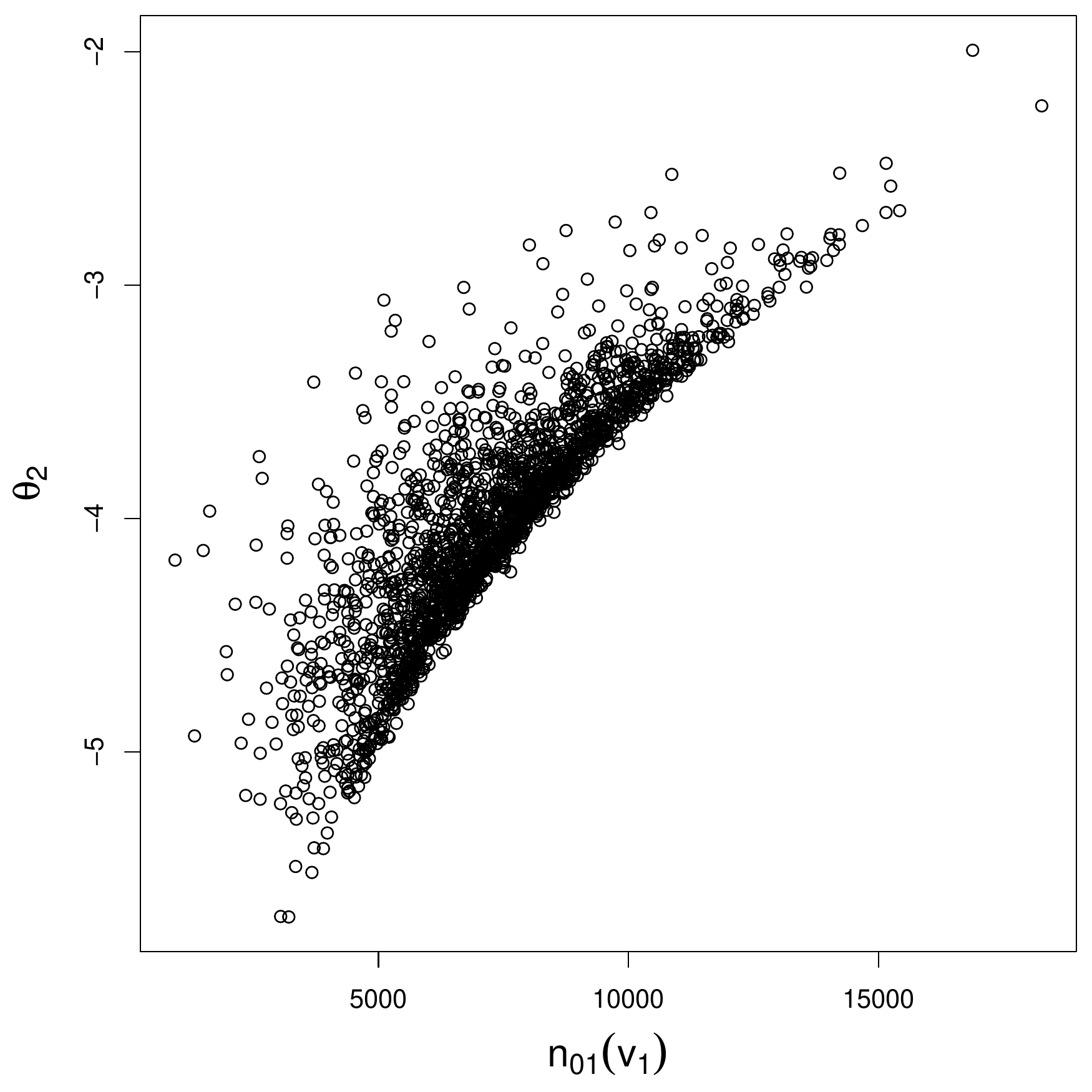} & 
\includegraphics[width=70mm]{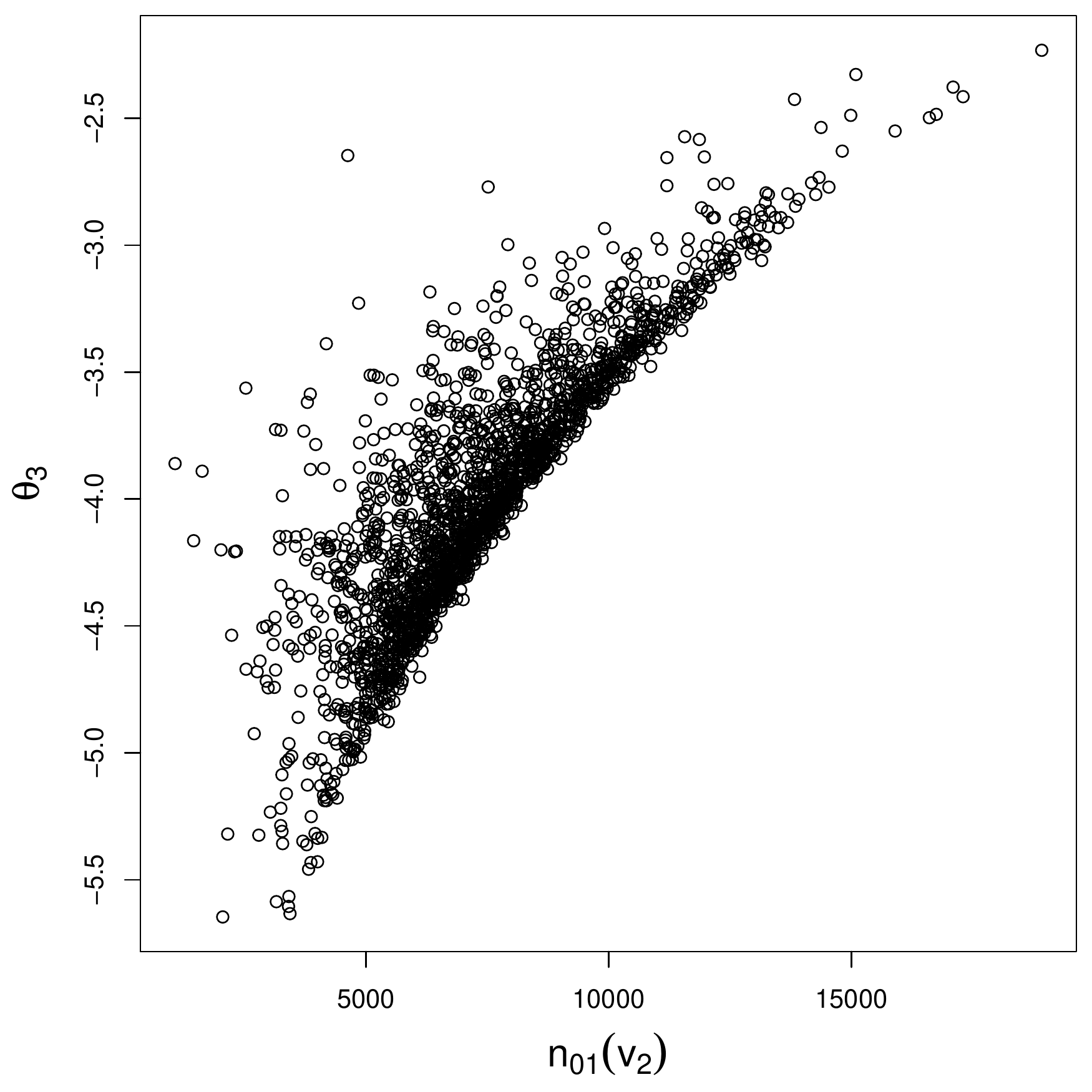} \\
\end{tabular}
\end{center}
\caption{\label{scatterplot} 
Plot of $\theta_2$ versus $n_{01}(v_1)$ (left) and $\theta_3$ versus $n_{01}(v_2)$ (right) for samples from
the prior for the heather incidence data.}
\end{figure}

It is interesting to understand why the global regression approach fails here.  Some insight
can be gained from Figure \ref{scatterplot}, which illustrates (prior predictive) scatter plots of $\theta_2$ versus $n_{01}(v_1)$
and $\theta_3$ versus $n_{01}(v_2)$.  The summary statistics $n_{01}(v_1)$ and $n_{01}(v_2)$
are those which are most informative about $\theta_2$ and $\theta_3$ respectively.  If we consider
regression of each of these parameters on the summary statistics, the graphs show
that not only the mean and variance, but also higher order properties, such as skewness of the response, 
appear to change as a function of the summary statistics. As such,  the heteroscedastic regression-adjustment based
on flexible estimation of the mean and variance does not work well here.  Making the regression local for
each marginal helps to overcome this problem.  
%

\subsection{Analysis of an AWBM computer model}

We now examine methods for the analysis of computer models,
where we aim to account for uncertainty in
high-dimensional forcing functions, assessment of model discrepancy and data rounding.  An approximate treatment
of this problem is interesting from a model assessment point of view, where
we want to judge whether the deficiencies of a computer model are
such that the model may be unfit  for some purpose. 

 A computer model can be regarded as a 
function $y=f(\eta)$ where $\eta$ are  model
inputs and $y$ is a vector of outputs.  
In modelling some particular physical system, observed data, $d$, is typically available that corresponds to some subset of the model outputs, $y$.  
%
%
%
%
The model inputs, $\eta$, can be of different types.  Here we only
make the distinction between model parameters, $\theta^*$, and forcing function inputs, $\omega$, so that $\eta=(\omega,\theta^*)$.  
Commonly, measurements of the forcing function inputs are available, and uncertainty
in these inputs (due to e.g. sampling and measurement errors) will be ignored in any analysis due to the high-dimensionality
involved.  An uncertainty analysis (involving an order of magnitude assessment of output
uncertainty due to forcing function uncertainty) will often be performed, rather than attempting
to include forcing function uncertainty directly in a calibration exercise (see, for
example, \shortciteNP{goldstein+sv10}, for an example of this in the context
of a hydrological model).  
See e.g. \shortciteN{craig+gss97}, \shortciteN{goldstein+r09}, \shortciteN{kennedy+o01} and \shortciteN{goldstein+sv10}  for further discussion of different aspects of computer models.

We now assume that $y=f(\eta)$ corresponds
to a prediction of  the observed data $d$ in the model
\begin{eqnarray}
  d & = & f(\eta)+g+e,  \label{model}
\end{eqnarray}
where $e$ denotes  measurement error and other sources of error independent in time, and $g$
is a correlated error term representing external model discrepancy (see \shortciteNP{goldstein+sv10} for a discussion of the differences between internal and external model discrepancies).
We directly investigate
forcing function uncertainty, through the term $f(\eta)$, using ABC. 
 In the analysis of the 
model (\ref{model}), we also consider data rounding effects, so that simulations produced
from (\ref{model}) are rounded according to the precision of the data that was
collected.  Handling such rounding effects is very simple in the ABC framework.

As a computer model, we consider
the Australian Water Balance Model (AWBM) (\citeNP{boughton04}), a rainfall-runoff model
widely used in Australia for applications such as
estimating catchment water yield or designing flood management systems.  
\begin{figure}
\begin{center}
\begin{tabular}{c}
\includegraphics[width=90mm]{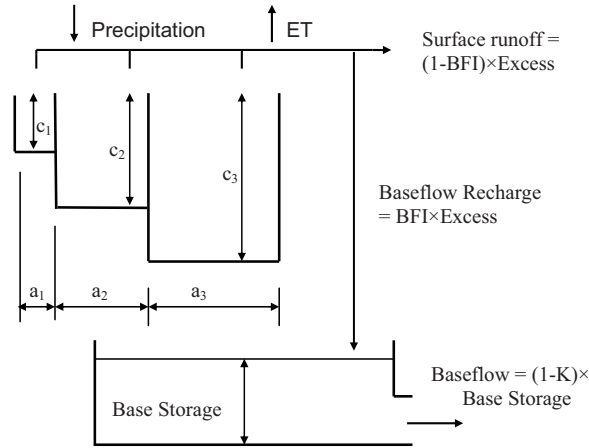}  \\ 
\end{tabular}
\end{center}
\caption{\label{awbm}  The three-catchment Australian water balance model.}
\end{figure}  
As shown in Figure \ref{awbm}, the model consists of three surface stores, with depths $c_1, c_2, c_3$ and
fractional areas $a_1, a_2, a_3$ with $\sum_{k} a_k=1$, and  a base store.
Model forcing inputs are precipitation and evapotranspiration time series, from which a predicted streamflow is produced.  
%
At each time step in the model, precipitation is  added to the system and evapotranspiration subtracted, with the net input 
split between the surface stores in proportion to the fractional areas.  
Any excess above the surface
store depths is then split between surface
runoff and flow into the base store according to the baseflow index $0<BFI<1$.  
Water from the base store is discharged into the stream at a rate determined by the recession
constant $0<K<1$, and the total discharge (streamflow) is then determined as the sum of the surface runoff and the baseflow.  
Following \citeN{bates+c01}, we fix $BFI=0.4$, although in some applications it may be
beneficial to allow this parameter to vary.  
The model parameters are therefore  $\theta^*=(c_1,c_2,c_3,a_1,a_2,K)$,
as well as the high-dimensional evapotranspiration and precipitation forcing inputs, $\omega$.
In hydrological applications there is often great uncertainty about the precipitation
inputs in particular, due to measurement and sampling errors.
Here we assume that evapotranspiration is fixed (known), and we use $\omega_{obs}$ to denote
the series of observed precipitation values only.   In running the computer model, we initialize with all stores empty and discard the first 500 days of the simulation to discount the effect of the assumed initial
conditions. 
Our data consist of a sequence of 5500 consecutive daily streamflow values from a station at Black River 
at Bruce Highway in Queensland, Australia.  The catchment covers an area of 260km$^2$ with a mean annual
rainfall of 1195mm.

To complete the determination of the computer model (\ref{model}), we specify the model priors.
Writing $\eta=(\omega,\theta^*)$, we describe the uncertainty on the true forcing inputs, $\omega=(\omega_1,\ldots,\omega_T)$, as
$\omega_t=\delta_t \omega_{obs,t}$, where the random multiplicative terms have prior $\log \delta_t\sim N(-\sigma_\delta^2/2,\sigma_\delta^2)$  for $t=1,...,T$.  
We set $\sigma_\delta\sim U(0,0.1)$, and note that $E(\delta_t)=1$ {\it a priori}.
%
For the external model discrepancy parameters, $g=(g_1,\ldots,g_T)^\top$, we specify  $g\sim N(0,\Sigma_g)$ 
where $\Sigma_g=[\Sigma_{g,ij}]$ is such that $\Sigma_{g,ij}=\sigma_g^2 \exp(-\rho |i-j|)$ , $\sigma_g\sim U(0,0.1)$ and $\rho\sim U(0.1,1)$.  
Independent model errors, $e=(e_1,\ldots,e_T)^\top$,  are assumed to be $e_t\sim N(0,\sigma_e^2)$, for $t=1,...,T$ where $\sigma_e\sim U[0,2]$.  
AWBM parameter priors are specified as
$c_1\sim U(0,500)$, $c_2-c_1\sim U(0,1000)$, $c_3-c_2\sim U(0,1000)$, 
$[\log(a_1/a_3),\log(a_2/a_3)]^\top\sim N(0,0.5^2 I)$ and $K\sim 0.271\times \mbox{Beta}(5.19,4.17)+(1-0.271)\times \mbox{Beta}(255,9.6)$, where the latter is a mixture of beta distributions.
See \shortciteN{bates+c01} for discussion of the background knowledge leading to
this prior choice.

If we treat the forcing inputs, $\omega$, as nuisance parameters, our parameter of interest is $\theta=(\theta^{*\top},\gamma^\top)^\top$, the set of AWBM model parameters, and 
$\gamma=(\sigma_e,\sigma_g,\sigma_q,\rho)^\top$,
those parameters specifying distributions of the stochastic terms in (\ref{model}).  
The ABC approach provides a convenient way of integrating out the high-dimensional nuisance parameter,
$\omega$, while dealing with complications such as rounding in the recorded data (the streamflow data are rounded to the nearest 0.01mm).  This would be very challenging
using conventional Bayesian computational approaches.

To define summary statistics, denote  $\hat{\theta}^*$ as the posterior mode estimate of $\theta^*$ 
in a model where we assume no input uncertainty, $\omega=\omega_{obs}$, and where we log-transform both 
the data and model output.  Also denote by $s_\gamma=[\psi(0),\psi(1),\psi(2),\zeta(1),\zeta(2),\zeta(3)]^\top$,
where $\psi(j)$ is the lag $j$ autocovariance of the least squares residuals $d-f(\hat{\eta})$, and $\zeta(j)$ is the
lag $j$ autocovariance of $(d-f(\hat{\eta}))^2$ with
$\hat{\eta}=(\omega_{obs},\hat{\theta}^*)$.  
In the notation of Section \ref{section:method}, for summary statistics for $\theta_j$, $j=1,...,6$ (i.e.
the components of $\theta^*$; the AWBM parameters) we use the statistic $s(j)=\hat{\theta}^*_j$ and for $\theta_j$, $j=7,...,10$
(i.e. the components of $\gamma$) we use the statistic $s(j)=s_\gamma$. 
 In effect, the
summary statistics for $\theta^*$ consist of point estimates for the AWBM parameters under the assumption of no error in the forcing inputs, $\omega$,
and statistics for the model error parameters, $\gamma$,
are intuitively based on autocovariances of residuals and squared residuals.  
Optimisation
of $\hat{\theta}^*$ is not trivial, as the objective function may have multiple modes.  To provide some degree of robustness, we select the best of ten Nelder-Mead simplex optimisations (\citeNP{nelder+m65}) using starting values simulated from the prior.
%

\begin{figure}
\begin{center}
\begin{tabular}{c}
\includegraphics[width=120mm]{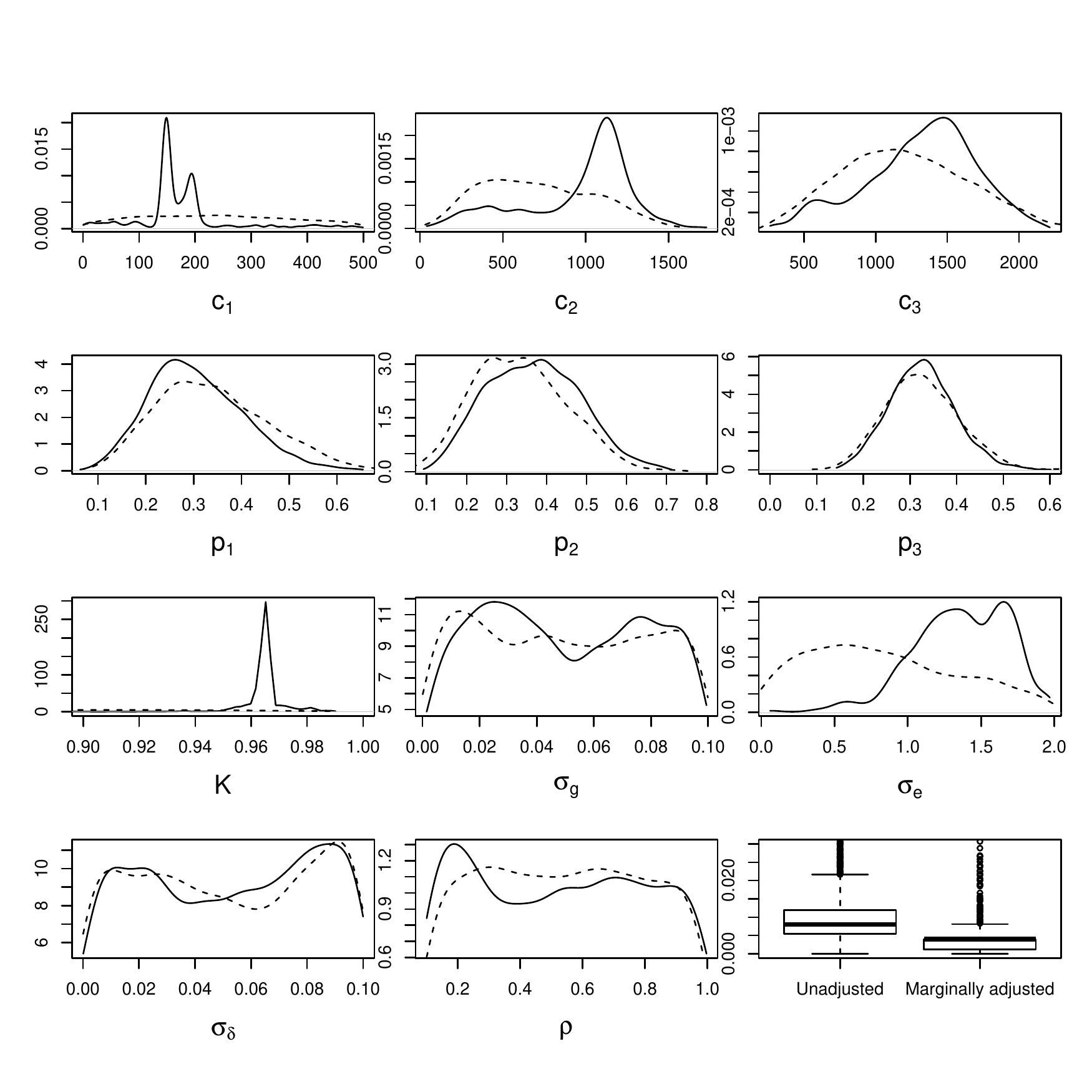}  \\ 
\end{tabular}
\end{center}
\caption{\label{awbmmarginals}  Estimated marginal posterior distributions for the AWBM computer model.  
Solid lines and dotted lines represent the seperately and jointly estimated marginals respectively.  The boxplots
on the bottom right show a within sample measure of fit of the AWBM model output over posterior samples
for the unadjusted and marginally adjusted methods.}
\end{figure}

Estimated marginal posterior distributions for the parameters are shown in Figure \ref{awbmmarginals}.  
\begin{figure}
\begin{center}
\begin{tabular}{ccc}
\includegraphics[width=50mm]{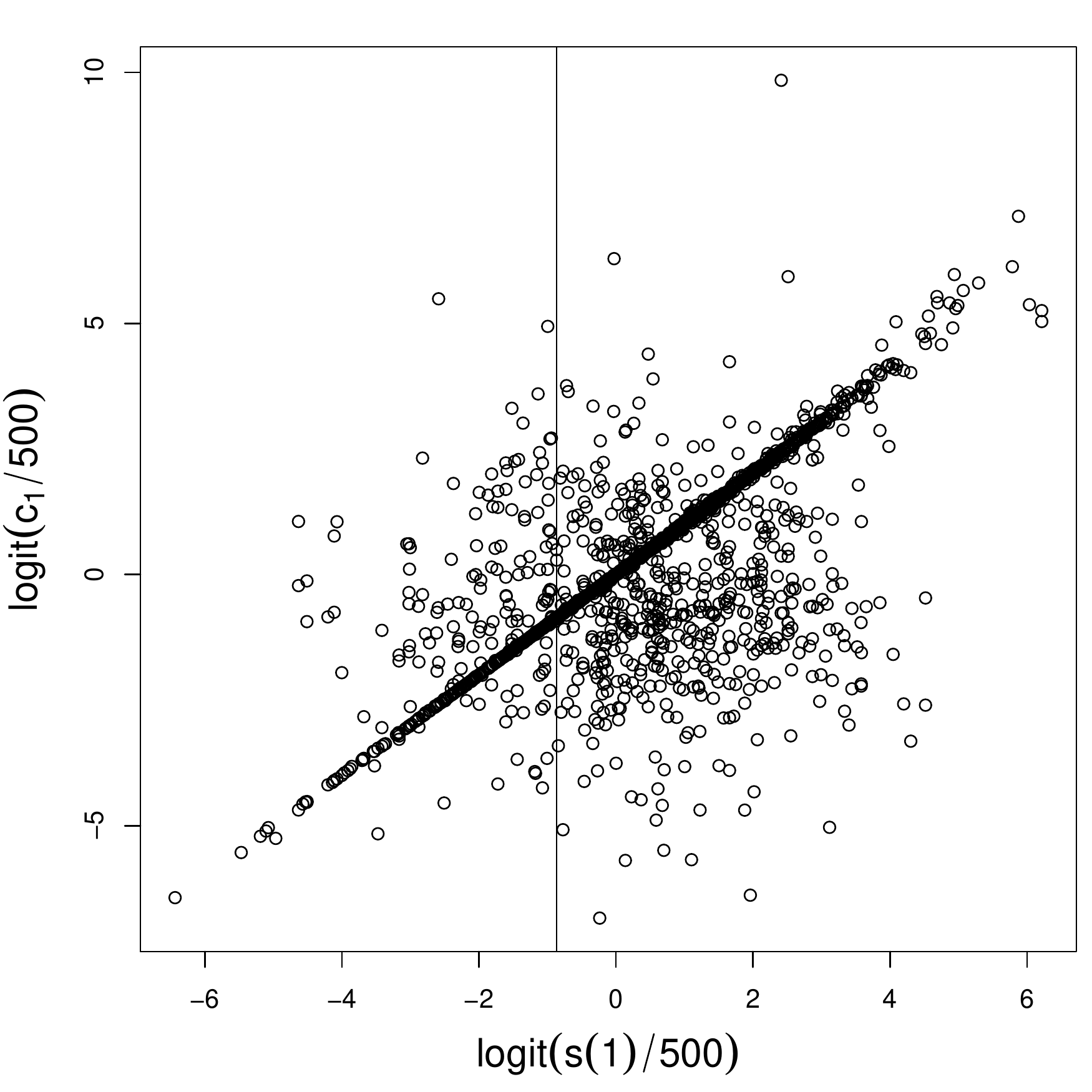} & 
\includegraphics[width=50mm]{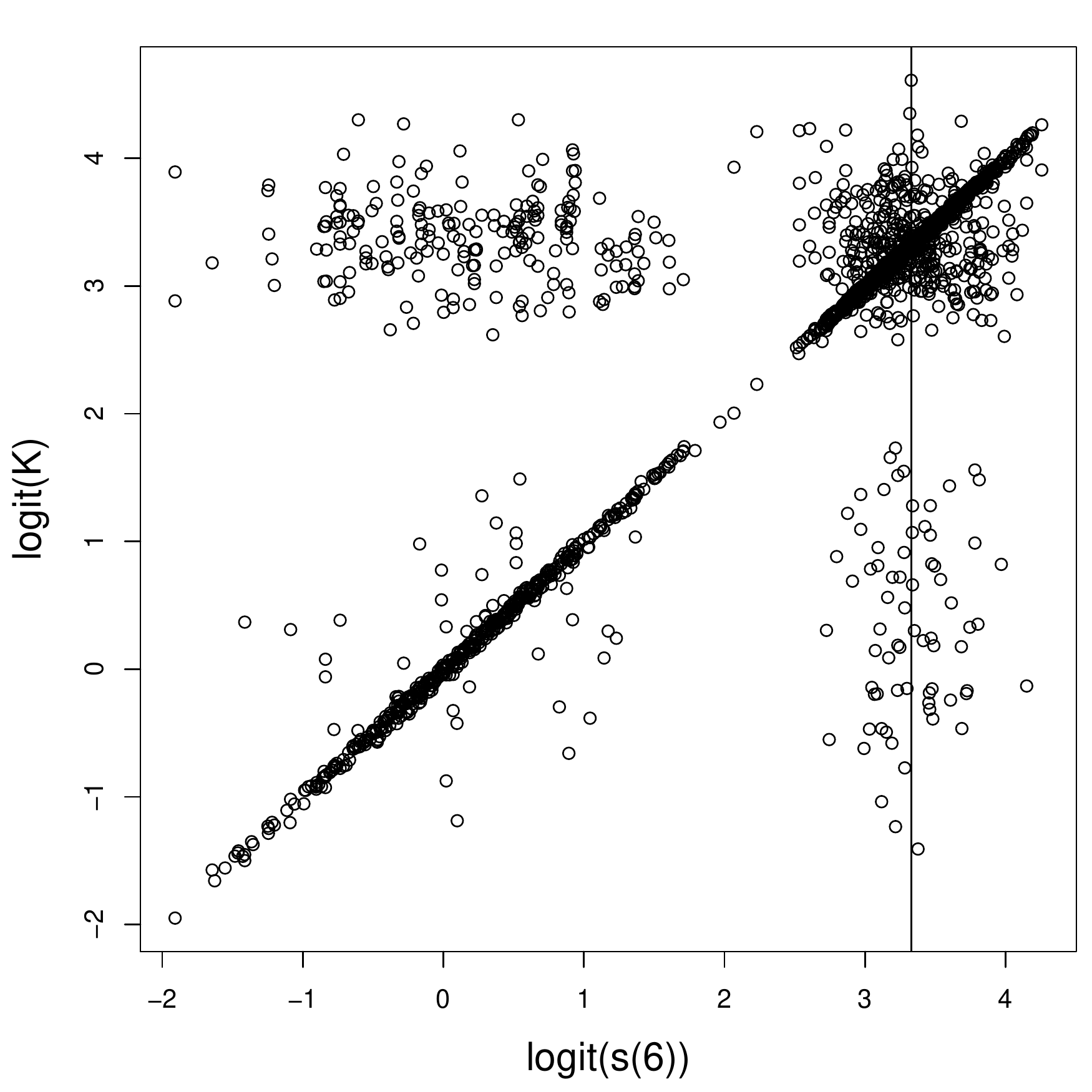}  & 
\includegraphics[width=50mm]{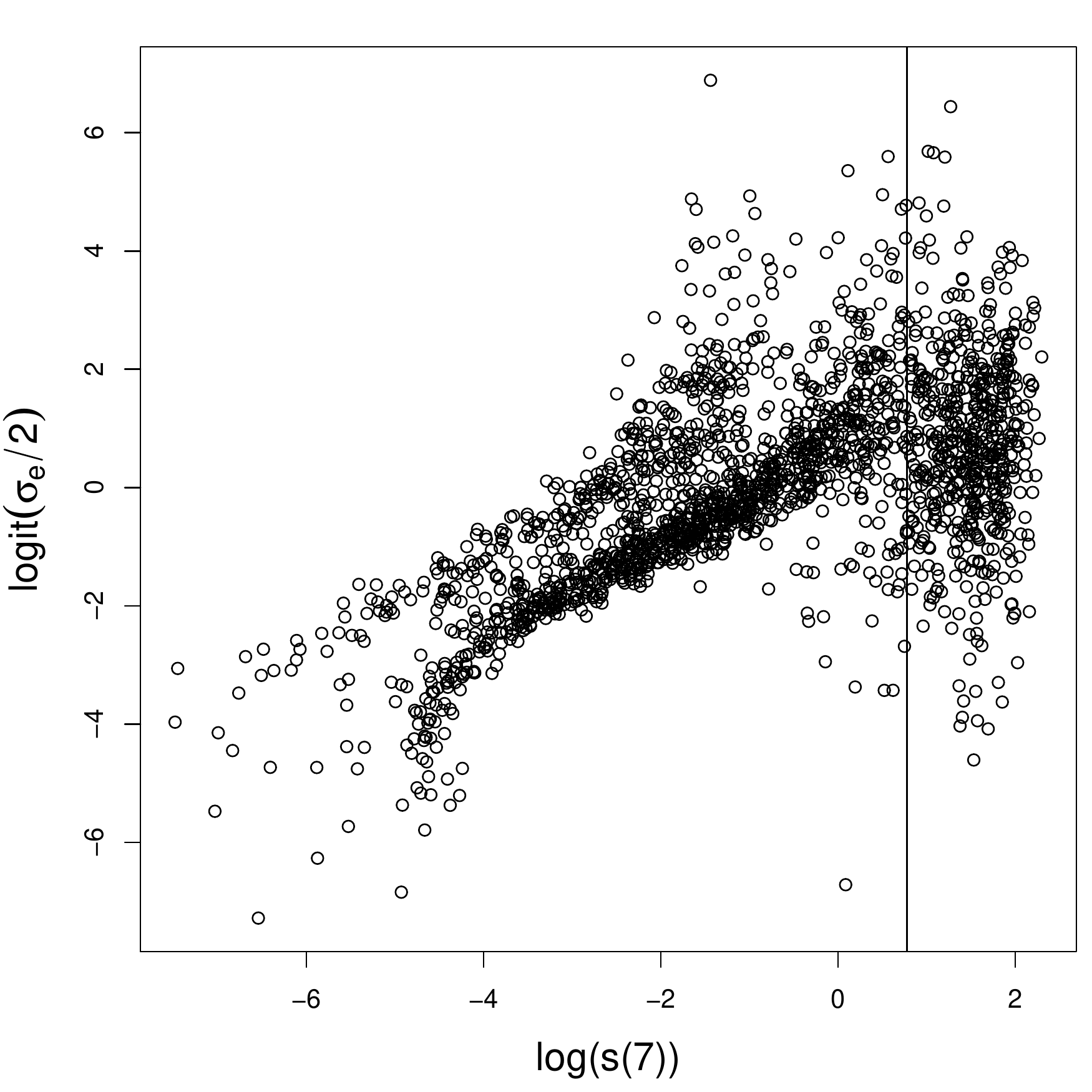}  \\
\end{tabular}
\end{center}
\caption{\label{awbmscatterplots} 
Plots of transformed parameters $c_1$, $K$ and $\sigma_e$ against transformed summary
statistics for samples from
the prior distribution for the AWBM example.  The solid vertical lines indicate the
observed values for the summary statistics.}
\end{figure} 
For the joint-posterior analysis, we implemented the non-linear, heteroscedastic, regression-adjustment of \citeN{blum+f10} 
using the uniform kernel, $K_\epsilon(\|\cdot\|)$, with scale parameter set to give non-zero weight to all 2,000 samples $(\theta^i,s^i)\sim p(s|\theta)p(\theta)$. For the individually estimated margins, the scale parameter was specified to select the 500 simulations closet to each $s(j)$. The discrepancy between
estimates for the parameters $c_1$, $K$ and $\sigma_e$ is particularly striking.  To understand why the
joint posterior regression-adjustment fails, 
Figure \ref{awbmscatterplots} shows prior predictive scatterplots of these parameters, each against their most informative summary statistic. 
Similar to the heather incidence example, the distribution of the response evidently changes as a function
of the covariates in more complicated ways than just through the first two moments. This is the root
cause of the difficulties with the joint regression-adjustment approach.  
Clearly,  the fact that the unadjusted
marginals are centred in the wrong place is unacceptable for inferential purposes.  

It is difficult here to do cross-validation or the usual
predictive checks since this would involve simulating from the posterior distribution for the high-dimensional
nuisance parameter, $\omega$, which is precisely what we have used ABC methods to avoid doing.  Instead, we examine the raw output of the AWBM model based on the available posterior, assuming no input uncertainty and model discrepancy. As a measure of model fit, we compute the median absolute error over time compared to the observed streamflow, on a log scale i.e. 
$\mbox{median}\left[\mbox{abs}(\log(d+1))-\log(f(\hat{\eta})+1))\right]$.
The boxplots in  Figure \ref{awbmmarginals} (bottom right panel)
show the distribution
of this within sample measure of fit across the two methods.  The highly inappropriate
posterior estimate of the unadjusted method leads to a worse model fit.  Although we have ignored model discrepancy and 
input uncertainty, we believe these results provide some independent verification of the unsuitability of 
the unadjusted posterior.  

A tentative conclusion from the above analysis is that input uncertainty (through the multiplicative
perturbation on the precipitation inputs, $\omega$, controlled through the term $\sigma_\delta$) may explain more of the model
misfit than the external model discrepancy term ($g$). As such, the AWBM
may be an acceptable model for the data given the inherent uncertainty in the forcing inputs.  

\section{Discussion}
\label{section:discussion}

In problems of moderate or high dimension, conventional sampler-based ABC methods which use 
rejection or importance-weight mechanisms, are of limited use.  As an alternative, regression-adjustment methods 
can be useful in such situations, however their accuracy as approximations to Bayesian inference 
may be difficult to validate.  

In this article we
have suggested that many regression-adjustment models are usefully viewed as Bayes 
linear approximations, which lends support to their utility in high dimensional ABC. We have also demonstrated
that it is possible to efficiently combine regression-adjustment methods with any ABC method (even sampler-based ones)
that can estimate a univariate marginal posterior distribution, in order to improve the quality of the 
ABC posterior approximation in higher dimensional problems.

In principle, our marginal-adjustment strategy can be applied to problems in any dimension. 
Given an initial sample from the joint ABC posterior sample, we propose to more precisely estimate and then replace its univariate marginal distributions. In terms of marginal precision, this idea is particularly viable using dimension reduction techniques which construct  one summary statistic per parameter (e.g. \citeNP{fernhead+p12}). 

However, this approach does not modify the dependence structure of the initial sample.  As such, if the dependence structure of the initial sample is poor (which can rapidly occur as the model dimension increases) then marginal-adjustment, however accurate, will not produce a fully accurate posterior approximation. 
Taken to the extreme, as the model dimension increases, the marginally-adjusted approximation will roughly constitute a product of independent univariate marginal estimates. While this is obviously less than perfect, a joint posterior estimate with independent, but well estimated margins, is a potential improvement over a very poorly estimated joint distribution.  Indeed, as pointed out by an anonymous referee, an expectation under the posterior that is linear in the parameters will still be well estimated.

In summary,  our marginal-adjustment strategy allows the application of standard ABC methods to problems of moderate to high dimensionality, which is comfortably beyond current ABC practice.
We believe that regression approaches in ABC
are likely to undergo further active development in the near future, as interest in ABC for more 
complex and higher dimensional models increases.  

\subsection*{Acknowledgements}

SAS is supported by the Australian Research Council through the Discovery Project Scheme (DP1092805).
The Authors thank M. G. B. Blum for useful discussion on regression-adjustment methods.
DJN gratefully acknowledges the support and contributions of the Singapore-Delft
Water Alliance (SDWA). The research presented in this work was carried out as part of
the SDWAs multi-reservoir research programme (R-264-001-001-272).
\bibliographystyle{chicago}
\bibliography{BLA}

\section{Supplementary Materials}

\begin{description} 

\item[Toy mixture-of-normals example:]

To reproduce the plots/analysis in the toy mixture-of-normals example, use the files
\begin{itemize}
\item make-mixture-plot.R which contains the R code, and
\item entropy-out-FULL.txt which contains the entropy estimates for one subplot.
\end{itemize}

\item[Excursion set analysis:]

To reproduce the heather excursion set analysis, use the files
\begin{itemize}
\item excursion.R which contains the R code, and
\item heather.txt which contains the heather data.
\end{itemize}

\item[AWBM analysis:]

To reproduce the AWBM analysis, use the files
\begin{itemize}
\item AWBM.R which contains the R code
\item sumstats.txt which contains the simulated summary statistics
\item sumstatsobs.txt which contains the observed summary statistics
\item params.txt which contains the simulated parameter values.
\end{itemize}

The R code may require the installation of additional libraries available on the CRAN.
\end{description}

\end{document}